\documentclass[12pt]{iopart}%

\def \BEA { \begin{eqnarray}}
\def \EEA {\end{eqnarray}}

\begin{document}

\title{Curvature tensors on distorted Killing horizons and their algebraic classification}
\author{V. Pravda\dag\ and O. B. Zaslavskii\ddag}
\date{\today}

\begin{abstract}
We consider generic static spacetimes with Killing horizons and study
properties of curvature tensors in the horizon limit. It is determined that
the Weyl, Ricci, Riemann and Einstein tensors are algebraically special and
mutually aligned on the horizon. It is also pointed out that results obtained
in the tetrad adjusted to a static observer in general differ from those
obtained in a free-falling frame. This is connected to the fact that a static
observer becomes null on the horizon.

It is also shown that finiteness of the Kretschmann scalar on the horizon is
compatible with the divergence of the Weyl component $\Psi_{3}$ or $\Psi_{4}$ in the freely
falling frame. Furthermore finiteness of $\Psi_{4}$ is compatible with
divergence of curvature invariants constructed from second derivatives of the
Riemann tensor.

We call the objects with finite Krestschmann scalar but infinite $\Psi_{4}$
``truly naked black holes''. In the (ultra)extremal versions of these objects
the structure of the Einstein tensor on the horizon changes due to extra terms
as compared to the usual horizons, the null energy condition being violated at
some portions of the horizon surface. The demand to rule out such divergencies
leads to the constancy of the factor that governs the leading term in the
asymptotics of the lapse function and in this sense represents a formal analog
of the zeroth law of mechanics of non-extremal black holes. In doing so, all
extra terms in the Einstein tensor automatically vanish.

\end{abstract}

\address{\dag Mathematical Institute, Academy of Sciences, \v{Z}itn\'{a} 25, 115 67 Prague
1, Czech Republic}
\address{\ddag Department of Mechanics and Mathematics, Kharkov V.N. Karazin's National
University, Svoboda Sq. 4, Kharkov 61077, Ukraine} \eads{\mailto{pravda@math.cas.cz}, \mailto{ozaslav@kharkov.ua}}

\section{Introduction}

The outstanding role of black holes in general relativity makes the
investigation of properties of event horizons especially important. Findings
and developments in this area till the seventies were summarized in
fundamental surveys \cite{Carter73}, \cite{Carter 79}. Nowadays, the interest
in this topic has been revived due to appearance of new notions such as
isolated and dynamical horizons (see \cite{ashliv} for review and references),
the issue of black hole entropy, including conformal field theory in the
near-horizon region \cite{conf}, etc.

Recently, it has been derived by purely geometrical means that for a generic
non-extremal static black hole the Einstein tensor has a high degree of
symmetry in the horizon limit \cite{med}. In the present paper, it is shown that for
a generic non-extremal static black hole the Weyl, Riemann, Ricci and Einstein
tensors are algebraically special and mutually aligned on the horizon and this
analysis is also extended to the extremal and ultraextremal cases.

Conditions for the regularity of the horizon are also studied. It turns out
that one can distinguish several classes by imposing various regularity
conditions. Throughout the paper the regularity condition from \cite{med},
that the Kretschmann invariant and all other polynomial invariants of the
Riemann tensor are regular on the horizon, is used. However it turns out that
even after imposing the above regularity condition the Weyl components
$\Psi_{3}$ and/or $\Psi_{4}$ in parallelly propagated frame may still diverge
on the horizon\footnote{For simplicity, the surface $N=0$ is refered to as a
horizon even in the case when a singularity is located there.} and thus a
parallelly propagated curvature singularity \cite{hawking} may be located
there. Even if both regularity conditions mentioned above (finiteness of  Kretschmann invariant, $\Psi_{3}$ and $\Psi_{4}$ ) are satisfied, there
are still cases for which curvature invariants constructed from second
derivatives of the Riemann tensor diverge on the horizon.

Up to now, only partial results concerning classification of the curvature
tensors on the horizon of a generic\footnote{We do not need to assume that
energy conditions or the Einstein equations are satisfied. We also have no
assumptions concerning topology of the horizon.} static black hole were
obtained. In the paper \cite{px} it was shown that on the horizon of an
axially symmetric black hole the Petrov type of the Weyl tensor is D. However,
this result was obtained in the static frame which becomes singular in the
horizon limit and thus is not suitable for this purpose\footnote{This has also
been noticed by M. Ortaggio (private communication).}. We consider also a
frame attached to a freely falling observer and point out that the horizon
limits of the Petrov type in both frames are, in general, different. In the
frame of a free-falling observer the resulting Petrov type is II or more
special. It was found in \cite{ash1}, \cite{ash2} that for a generic isolated
horizon the Petrov type is II or more special, but without further specification.

Since for physical reasons, physical properties of the black holes are often
analyzed in the static frame, which is regular everywhere in the outer region
but becomes null on the horizon, we also perform calculations in the static
frame and compare them with results obtained in the frame attached to the
freely falling observer.

The paper is organized as follows. In Sec. \ref{sec2} we describe the choice
of the null tetrad and, using 2+1+1 decomposition of the metric and curvature,
list basic formulas for the Weyl scalars $\Psi_{0},\ \dots,\ \Psi_{4}$. We
apply them to the static frame and consider separately non-extremal, extremal
and ultraextremal horizons. In Sec. \ref{sec3} we consider a free-falling
frame. As explicit examples, we discuss the Ernst metric describing a
Schwarzschild-like black hole in a magnetic field and the Bonnor-Swaminarayan
metric. In Sec. \ref{sec4} we show that, in general, a free-falling observer
can register divergence of the Weyl scalar $\Psi_{4}$ near the horizon in
spite of finiteness of the Kretschmann invariant. We discuss conditions that
rule out such objects that we call ``truly naked black holes''. In Sec.
\ref{sec5} we determine an algebraic type of the Ricci tensor on the horizon.
Sec. \ref{sec6} contains a list of main results and their brief discussion.

\section{Static observer}

\label{sec2}

Let us consider a generic static spacetime. We follow the technique of
decomposition of the metric and curvature \cite{israel}, \cite{med}. The
metric can be written in the 3+1 decomposed form with a subsequent 2+1
decomposition on the basis of the Gauss normal coordinates

\begin{equation}
ds^{2}=-dt^{2}N^{2}+dn^{2}+\gamma_{ab}dx^{a}dx^{b}{,}\label{m}%
\end{equation}
where $x^{1}=n$, $a=2,3$. { As our spacetime is static, there exists a
Killing vector which is time-like in the outer region, }$\xi^{\mu}%
=(1,0,0,0)$, {  its norm }$\xi^{\mu}\xi_{\mu}\equiv\xi^{2}=-N^{2}%
<0$. {  We also suppose that there exists a two-dimensional regular
two-surface obtained by the limiting transition }$N^{2}=c$,
$c\rightarrow+0$ { (}$c$ {  is a constant) on which the Killing
vector becomes null, }$\xi^{2}=0$. { This surface separates the outer
region in which this vector is time-like from that where it is spacelike
(}$\xi^{2}>0$). { In other words, this is the so-called Killing horizon.
It acts also as a surface of an infinite redshift. Its significance in black
hole context is connected, in particular, with the fact that for any black
hole solution in the stationary (in our case static) spacetime, with matter
satisfiying suitable hyperbolic equations, the event horizon is a Killing
horizon \cite{hawking} (Prop. 9.3.6), \cite{nf} (Sec. 6.3.1). A rigorous
definition of a black hole usually assumes asymptotical flatness. Then a black
hole is a region from which no causal signal can reach future null infinity
(see, e.g. Sec. 5.2.1 of \cite{nf}). However, we would like to stress, that
for our purposes we do not distinguish black hole horizons and acceleration
horizons (but we speak sometime about black hole for the sake of simplicity
and definiteness), do not require asymptotic flatness, etc. All we need is the
local properties of the metric and curvature tensors that follow from the fact
that }$N^{2}=0$ {  and the regularity of spacetime (see details below).}

{ Our goal is to elucidate, to what extent the presence of the Killing
horizon restricts the Petrov type of the gravitational field on the horizon,
and find which types are possible there. Our
determination of the Petrov type is based on studying invariants constructed
from so-called Weyl scalars (see their exact definition below):}%

\begin{equation}
\mathrm{I=\Psi}_{0}\mathrm{\Psi}_{4}\mathrm{-4\Psi}_{1}\mathrm{\Psi}%
_{3}\mathrm{+3\Psi}_{2}^{2}\mathrm{,\ J=}\det\left(
\begin{array}
[c]{ccc}%
\Psi_{4} & \Psi_{3} & \Psi_{2}\\
\Psi_{3} & \Psi_{2} & \Psi_{1}\\
\Psi_{2} & \Psi_{1} & \Psi_{0}%
\end{array}
\right)   {,} \label{I}%
\end{equation}
$\mathrm{\\}$%
\begin{equation}
\mathrm{K=\Psi}_{1}\mathrm{\Psi}_{4}^{2}\mathrm{-3\Psi}_{2}\mathrm{\Psi}%
_{3}\mathrm{\Psi}_{4}\mathrm{+2\Psi}_{3}^{3}\mathrm{,\ L=\Psi}_{2}%
\mathrm{\Psi}_{4}\mathrm{-\Psi}_{3}^{2}\mathrm{,\ N=12L}^{2}\mathrm{-\Psi}%
_{4}^{2}\mathrm{I.} \label{K}%
\end{equation}

{ The algorithm for determining Petrov type of the Weyl tensor is based
on whether or not equalities }%
\begin{equation}
I^{3}=27J^{2}\mathrm{,\ }I=J=0\mathrm{,\ }K=N=0\mathrm{,\ }K=L=0 \label{crit}%
\end{equation}
{  are satisfied (see, e.g., \cite{kram}). Our strategy can be described
as follows.}

1) We choose the complex tetrad frame and, with its help, define Weyl
scalars,

2) we use 2+1+1 splitting of the metric of the static spacetime using
Gauss normal coordinates and find general expressions for Weyl scalars,

3) the conditions of the regularity of spacetime on the horizon impose
severe restrictions on the asymptotic form of the metric, we substitute this
asymptotics in the formulas for Weyl scalars and find their near-horizon
values,

4) compare the result with the conditions that define the Petrov
type,

5) carry out this procedure for non-extremal and (ultra)extremal
horizons separately,

6) repeat it for a tetrad that corresponds to a free-falling
observer.

Let us construct the complex null tetrad from a usual orthonormal frame
$u^{\mu}$, $e^{\mu}$, $a^{\mu}$, $b^{\mu}$, where $u^{\mu}$ is a four-velocity
of an observer, $e^{\mu}$ is a vector aligned along the $n$-direction,
$a^{\mu}$ and $b^{\mu}$ lie in the $x^{2}-x^{3}$ subspace. We define%

\begin{equation}
l^{\mu}=\frac{u^{\mu}+e^{\mu}}{\sqrt{2}}{,}\ n^{\mu}=\frac{u^{\mu}-e^{\mu}%
}{\sqrt{2}},\ m^{\mu}=\frac{a^{\mu}+ib^{\mu}}{\sqrt{2}},\ \bar{m}^{\mu}%
=\frac{a^{\mu}-ib^{\mu}}{\sqrt{2}}{.}%
\end{equation}
Now $l^{\mu}n_{\mu}=-1$, $m^{a}\bar{m}_{a}=1$, all other contractions vanish.
We use the standard definition of the Weyl scalars 
\BEA
\fl
\Psi_0=C_\alpha\beta\gamma\delta l^\alpha m^\beta l^\gamma m^\delta,
\ \Psi_1=C_\alpha\beta\gamma\delta l^\alpha m^\beta l^\gamma n^\delta,
\ \Psi_2=-C_\alpha\beta\gamma\delta l^\alpha m^\beta n^\gamma\bar{m}^\delta,
\label{psi02} \nonumber \\
\Psi_3=C_\alpha\beta\gamma\delta l^\alpha
n^\beta\bar{m}^\gamma n^\delta, \ \Psi_4=C_\alpha\beta\gamma\delta
n^\alpha\bar{m}^\beta n^\gamma\bar{m}^\delta{.} \label{psi34}\EEA
Here the Weyl
tensor%
\begin{equation}
C_{\alpha\beta\gamma\delta}=R_{\alpha\beta\gamma\delta}-R_{\gamma\lbrack
\alpha}g_{\beta]\delta}+R_{\delta\lbrack\alpha}g_{\beta]\gamma}+\frac{R}%
{3}g_{\gamma\lbrack\alpha}g_{\beta]\delta} {,}\label{c}%
\end{equation}
where $R_{\alpha\beta\gamma\delta}$ is the curvature tensor, $R$ is the scalar
curvature. Our first goal is to find behavior of these scalars near the
horizon whence we will be able to extract information about the Petrov type.
In what follows we will be dealing with two types of such tetrads
corresponding to a static observer (SO) and a freely falling one (FFO).

In the present section, we consider the first case (SO). Then
\begin{equation}
u^{(0)\mu}=(N^{-1},0,0,0),\ \ e^{(0)\mu}=(0,1,0,0){,}\label{ues}%
\end{equation}
where the superscript (0) refers to the static frame, so that%
\begin{equation}
l^{\mu}=\frac{1}{\sqrt{2}}(\frac{1}{N},1,0,0),\ \ n^{\mu}=\frac{1}{\sqrt{2}%
}(\frac{1}{N},-1,0,0),\ \ m^{\mu}=(0,0,m^{a}){.}\label{ln}%
\end{equation}
With this choice of the tetrad,
\begin{equation}
\Psi_{4}=\bar{\Psi}_{0},\Psi_{3}=-\bar{\Psi}_{1}{.}\label{01}%
\end{equation}
Thus, it is sufficient to determine $\Psi_{0}$, $\Psi_{1}$, $\Psi_{2}$.

{ In what follows we will extensively use the convenient representation
of the curvature tensor based on 2+1+1 decomposition of the metric (\ref{m})
(details can be found in Sec. 2 of Ref. \cite{med}):}%
\begin{equation}
R_{abcd}=\frac{R_{\parallel}}{2}(\gamma_{ac}\gamma_{bd}-\gamma_{ad}\gamma
_{bc})+K_{ad}K_{bc}-K_{ac}K_{bd}\label{abcd}%
\end{equation}%
\begin{equation}
R_{1abc}=K_{ac;b}-K_{ab;c}\label{1abc}%
\end{equation}%
\begin{equation}
R_{1a1b}=\frac{\partial K_{ab}}{\partial n}+\left(  K^{2}\right)
_{ab}\label{1a1b}%
\end{equation}%
\begin{equation}
\frac{R_{0a0b}}{N^{2}}=\frac{N_{;a;b}-K_{ab}N^{\prime}}{N}\label{0a0b}%
\end{equation}%
\begin{equation}
\frac{R_{0101}}{N^{2}}=\frac{N^{\prime\prime}}{N}=-\frac{1}{2}R_{\perp
} {,}\label{0101}%
\end{equation}%
\begin{equation}
\frac{R_{010a}}{N^{2}}=\frac{\partial_{n}N_{;a}+K_{a}^{b}N_{;b}}{N}%
 {,}\label{010a}%
\end{equation}
{ where }$K_{ab}${  is the extrinsic curvature tensor for the
surface }$t=const$, {  }$n=\nolinebreak const$ {  embedded in the
outer three-space, for our metric (\ref{m}) }%
\begin{equation}
K_{ab}=-\frac{1}{2}\frac{\partial\gamma_{ab}}{\partial n},\label{kab}%
\end{equation}
$K=K_{a}^{a}$, prime denotes differentiation with respect to
$n$, quantities like  $N_{\mid i\mid j}$  represent covariant
derivatives with respect to the three-metric and $N_{;b;d}$ { 
correspond to covariant derivatives with respect to the two-metric }%
$\gamma_{ab}$, $R_{\parallel}${  represents the two-dimensional
Ricci scalar for the surface }$t=const$, $n=const$, $R_{\perp}$ {  is the similar quantity for the }$n-t${  subspace.}

{ In a similar way, for the Ricci tensor one has}%
\begin{equation}
\frac{R_{00}}{N^{2}}=\frac{\Delta_{2}N-KN^{\prime}+N^{\prime\prime}}%
{N} {,}\label{00}%
\end{equation}%
\begin{equation}
R_{11}=K^{\prime}-SpK^{2}-\frac{N^{\prime\prime}}{N} {,}\label{11}%
\end{equation}%
\begin{equation}
R_{1a}=K_{;a}-K_{a}^{b}{}_{;b}-\frac{\partial_{n}N_{;a}+K_{a}^{b}N_{;b}}%
{N} {,}\label{r1a}%
\end{equation}

\begin{equation}
R_{ab}=\frac{R_{\parallel}}{2}\gamma_{ab}+\frac{\partial K_{ab}}{\partial
n}+2(K^{2})_{ab}-KK_{ab}-\frac{N_{;a;b}-K_{ab}N^{\prime}}{N} {,}\label{ab}%
\end{equation}
{ where }$\Delta_{2}$ {  is the two-dimensional Laplacian with
respect to the metric }$\gamma_{ab}$. {  The scalar curvature}%
\begin{equation}
R=R_{\parallel}+2K^{\prime}-SpK^{2}-K^{2}-2\frac{\Delta_{2}N-KN^{\prime
}+N^{\prime\prime}}{N} {.}\label{R}%
\end{equation}

Now eq. (\ref{psi02}) leads to
\begin{equation}
\Psi_{0}=\frac{1}{2}m^{b}m^{d}\left(  \frac{C_{0b0d}}{N^{2}}+C_{1b1d}\right)
{,}%
\end{equation}
where $C_{0b0d}$ and $C_{1b1d}$ may be obtained from the definition of the
Weyl tensor (\ref{c}):%
\begin{equation}
C_{0b0d}=R_{0b0d}-\frac{R_{00}}{2}\gamma_{bd}+\frac{N^{2}}{2}R_{bd}-\frac
{R}{6}N^{2}\gamma_{bd}{,}%
\end{equation}%
\begin{equation}
C_{1b1d}=R_{1b1d}-\frac{R_{11}}{2}\gamma_{bd}-\frac{R_{bd}}{2}+\frac{R}%
{6}\gamma_{bd}{.}%
\end{equation}
Since $\gamma_{ab}m^{a}m^{b}=0$,
\begin{equation}
\Psi_{0}=\frac{1}{2}m^{b}m^{d}{\tilde{S}}_{bd},\ \ \mbox{where}\ \ {\tilde{S}%
}_{bd}=\frac{R_{0b0d}}{N^{2}}+R_{1b1d}{.}%
\end{equation}
Consequently
\begin{equation}
\Psi_{0}=\frac{1}{2}m^{b}m^{d}\left[  \frac{\partial K_{bd}}{\partial
n}-K_{bd}\frac{N^{\prime}}{N}+\left(  K^{2}\right)  _{bd}+\frac{N_{;b;d}}%
{N}\right]  .\label{0}%
\end{equation}
We also have%
\begin{equation}
\Psi_{1}=-\frac{m^{b}}{\sqrt{2}}\frac{C_{0b01}}{N^{2}}{,}\label{1}%
\end{equation}
where%
\begin{equation}
\frac{C_{0b01}}{N^{2}}=\frac{R_{0b01}}{N^{2}}+\frac{1}{2}R_{1b} {,}%
\end{equation}
whence%
\begin{equation}
\Psi_{1}=-\frac{m^{b}}{2\sqrt{2}}\left(  N^{-1}\frac{\partial N_{;b}}{\partial
n}+\frac{K_{b}^{c}N_{;c}}{N}+K_{;b}-\gamma^{cd}K_{bc;d}\right)  {.}\label{c1}%
\end{equation}
One also finds
\begin{equation}
-2\Psi_{2}=m^{b}\bar{m}^{d}\left[  \frac{R_{\parallel}}{2}\gamma_{bd}+\left(
K^{2}\right)  _{bd}-K_{bd}K\right]  +\frac{R_{11}}{2}-\frac{R}{3}-\frac
{R_{00}}{2N^{2}}{.}%
\end{equation}
After expressing this quantity in terms of two-dimensional geometry we obtain%
\begin{equation}
\fl-2\Psi_{2}=\frac{R_{\parallel}+R_{\perp}}{6}+\frac{K^{2}}{3}-\frac{SpK^{2}%
}{6}-\frac{K^{\prime}}{6}-\frac{KN^{\prime}}{6N}+\frac{\Delta_{2}N}{6N}%
+m^{b}\bar{m}^{d}[\left(  K^{2}\right)  _{bd}-K_{bd}K] {{.}}\label{2}%
\end{equation}
Now different types of horizons should be considered separately.

\subsection{Non-extremal case}

{ The necessary condition of absence of singularities is the finiteness
of the Kretschmann invariant}%
\begin{equation}
Kr\equiv R_{\alpha\beta\gamma\delta}R^{\alpha\beta\gamma\delta} {.}%
\end{equation}
{ For the static spacetimes (\ref{m}) it is easy to show \cite{israel},
\cite{med} that }$Kr={}^{(3)}R_{ijkl} {}^{(3)}R^{ijkl}+4\frac{N_{\mid i\mid
j}N^{N_{\mid i\mid j}}}{N^{2}}$. The first term here is calculated
with respect to the space positively defined three-metric and is positive, if
we want to have a regular spacetime, the second term (which is also positive)
should be finite. Then, if the surface gravity $\kappa_{H}\neq0$ { 
(}$\kappa_{H}$ {  is equal to the horizon limit of the derivative }%
$\frac{\partial N}{\partial n}$), one derives the general form of the
asymptotic behavior of the lapse function for small $n$   \cite{med}:%

\begin{equation}
N=\kappa_{H}n+\frac{\kappa_{2}(x,y)}{3!}n^{3}+\frac{\kappa_{3}(x,y)}{4!}%
n^{4}+O(n^{5}) {,} \label{nn}%
\end{equation}%
\begin{equation}
\gamma_{ab}=[\gamma_{H}]_{ab}(x,y)+\frac{[\gamma_{2}]_{ab}(x,y)}{2!}%
n^{2}+\frac{[\gamma_{3}]_{ab}(x,y)}{3!}n^{3}+O(n^{4}), \label{expgamma}%
\end{equation}
where $\kappa_{H}$ is a constant on the horizon. From these expansions we
obtain that in general near the horizon
\begin{equation}
\Psi_{0} {, }\ \Psi_{1} {, }\ \Psi_{3} {, }\ \Psi_{4}\ \sim\ n {.}%
\end{equation}
In the particular case with $[\gamma_{3}]_{ab}(x,y)=0$ which will be clarified
later we obtain
\begin{equation}
\Psi_{0},\ \Psi_{4} \ \sim n^{2},\ \Psi_{1}, \ \Psi_{3} \ \sim\ n.
\end{equation}
It also follows from (\ref{2}) that the horizon value of $\Psi_{2}$ is equal
to%
\begin{equation}
\Psi_{2}^{H}=-\frac{R_{\parallel}^{H}+R_{\perp}^{H}}{12} {.} \label{2h}%
\end{equation}

Thus, there exist only two possibilities on the horizon: (i) $\Psi_{0}$,
$\Psi_{1}$, $\Psi_{3}$, $\Psi_{4}$ vanish, $\Psi_{2}\neq0$, (ii) all
components of the Weyl tensor vanish. The case (i) corresponds to the Petrov
type D and the case (ii) to the Petrov type O. {We must make a reservation
here. As the static frame becomes singular on the horizon (we discuss it in
more detail below), in this section, by the Petrov type on the horizon we
simply mean the type obtained by taking the horizon limit from the outer
region.}

As an example of the case (ii), we can mention the Bertotti-Robinson (BR)
metric for which $R_{\parallel}+R_{\perp}=0$ everywhere and so it is of type
O. However, quantum backreaction of massless conformally invariant fields on
spacetimes of the type AdS$_{2} \times$S$_{2}$ (which the BR metric belongs
to) violates this condition \cite{accel}. By contrary, bacreaction of massive
fields retains its validity \cite{mz}. Thus, as far as the role of quantum
backreaction is concerned, conformal fields change the Petrov type of the
metric on the horizon from O to D, whereas massive fields leave it intact.

\subsection{Extremal case}

Let $\kappa_{H}=0$. In this case the horizon is situated at the infinite
proper distance from any point. In Ref. \cite{med} this was shown for the
function $N$ that has the power-like asymptotics at large $n$. This
corresponds to what is usually called \textquotedblleft
ultraextremal\textquotedblright\ horizon in the spherically-symmetrical case
(see, e.g. \cite{brill}), when $N^{2}\sim(r-r_{+})^{m}$, $m\geq3\,$, $r$ is
the Schwarzschild-like coordinate, $r_{+}$ is the position of the horizon. For
the case of the usual extremal black hole $m=2.$ The extremal
Reissner-Nordstr\"{o}m metric can serve as example, then $N\sim(r-r_{+}%
)\sim\exp(-\frac{n}{r_{+}})$. Thus, there are two qualitatively different
cases of the asymptotic behavior of the lapse function $N$. The similar
division occurs also in a general non-spherical case. It turns out that
finiteness of the Kretschmann invariant on the horizon requires that
$\frac{N_{\mid i\mid j}}{N}$ be finite there \cite{israel}, $i,j=1,2,3$. It
makes sense to distinguish two cases: (i) $\frac{N_{\mid1\mid1}}{N}%
=C(x^{2},x^{3})\neq0$ on the horizon, (ii) $\frac{N_{\mid1\mid1}}{N}=0$ on the horizon.

In the case (i) we obtain the asymptotic behaviour%
\begin{equation}
N=B(x^{2},x^{3})\exp(-\frac{n}{n_{0}})+O(e^{-\frac{2n}{n_{0}}}){,}\ n_{0}%
>0{,}\ \frac{1}{n_{0}^{2}}=C \label{e}%
\end{equation}
for $n\rightarrow\infty$. Then the dependence $n_{0}(x^{2},x^{3})$ would
produce in $\frac{N_{\mid b\mid d}}{N}=\frac{N_{;b;d}}{N}-K_{bd}%
\frac{N^{\prime}}{N}$ terms proportional to $n^{2}$ that diverge on the
horizon. Therefore, $n_{0}$ should be constant. Expanding the two-dimensional
metric and the extrinsic curvature tensor%
\begin{equation}
\fl \gamma_{ab}=\gamma_{ab}^{(0)}+\gamma_{ab}^{(1)}\exp(-\frac{n}{n_{0}})
+O(e^{-\frac{2n}{n_{0}}}){,}\ \ K_{ab}=K_{ab}^{(1)}\exp(-\frac{n}{n_{0}})
+O(e^{-\frac{2n}{n_{0}}})
\end{equation}
and calculating the Weyl scalars, we obtain that
\begin{equation}
\Psi_{0}^{H}=\frac{B_{;b;d}}{2B}m^{b}m^{d}{,}\ \ \Psi_{1}^{H}=\frac{m^{b}%
}{2n_{0}\sqrt{2}}\frac{B_{;b}}{B}{,} \label{0H}%
\end{equation}%
\begin{equation}
\Psi_{2}^{H}=-\frac{1}{12}(R_{\parallel}+R_{\perp}+\frac{\Delta_{2}B}{B}){,}
\label{2horb}%
\end{equation}
where the right hand sides are evaluated on the horizon.

\subsection{Ultraextremal case}

In the case (ii)%

\begin{equation}
N=\frac{A}{n^{m}}+O(n^{-m-1}){,}\ \ n\rightarrow\infty{,}\ \ m>0 \label{ue}%
\end{equation}%
\begin{equation}
\gamma_{ab}=\gamma_{ab}^{(0)}+\frac{\gamma_{ab}^{(1)}}{n^{s}}+O(n^{-s-1}%
){,}\ \ s>0
\end{equation}%
\begin{equation}
K_{ab}=\frac{K_{ab}^{(1)}}{n^{s+1}}+O(n^{-s-2}), \label{kabu}%
\end{equation}%
\begin{equation}
\Psi_{0}^{H}=\frac{A_{;b;d}}{2A}m^{b}m^{d}{,}\ \ \Psi_{1}^{H}=0{,}%
\end{equation}%
\begin{equation}
\Psi_{2}^{H}=-\frac{1}{12}(R_{\parallel}^{H}+R_{\perp}^{H}+\frac{\Delta_{2}%
A}{A}){.} \label{2hor}%
\end{equation}

We postpone discussion of the meaning of the terms with derivatives like
$B_{;b}$ or $A_{;a}$ and now pass to the description of the gravitational
field from the viewpoint of the FFO.

\section{Freely falling observer}

\label{sec3}

Let $u^{\mu}$ and $e^{\mu}$ denote vectors attached to a FFO that moves toward
the horizon. Then it follows from integration of equations of timelike
geodesics that for a \textquotedblleft radial\textquotedblright\ motion
$x^{2}=$ const, $x^{3}=$ const 
\BEA
\fl l^\mu=\left(  \frac{E}{N^2}-\frac
{\sqrt{E^2-N^2}}{N^2},\frac{E}{N}-\frac{\sqrt{E^2-N^2}}{N},0,0\right)
=e^{-\alpha}\left(  \frac{1}{N},1,0,0\right)  {, }\\\fl n^\mu=\left(
\frac{E}{N^2}+\frac{\sqrt{E^2-N^2}}{N^2},-\frac{E}{N}-\frac{\sqrt{E^2-N^2}}%
{N},0,0\right)  =e^\alpha\left(  \frac{1}{N},-1,0,0\right)  {,} \\%
\cosh\alpha=\frac{E}{N} {, }\ \alpha> 0 {,} 
\EEA
with $E=-u_{0}$ being the energy per unit mass.

Two pairs of vectors corresponding to SO and FFO are connected by the
relationships 
\BEA
u^{\mu}=u^{\mu(0)}\cosh\alpha-e^{\mu(0)}\sinh\alpha {,} \label{ua} \\
e^{\mu}=e^{\mu(0)}\cosh\alpha-u^{\mu(0)}\sinh\alpha {,} \label{ea} \\
l^{\mu}=zl^{(0)\mu} {, }\ n^{\mu}=z^{-1}n^{(0)\mu} {, }\ \ \mbox{with}\ \ z=\exp
(-\alpha) {.} \label{lna}%
\EEA
Under the boost (\ref{lna}) the Weyl scalars transform
in the standard way:%
\begin{equation}
\fl\Psi_{0}=z^{2}\Psi_{0}^{(0)},\ \Psi_{1}=z\Psi_{1}^{(0)}{,}\ \Psi_{2}%
=\Psi_{2}^{(0)},\ \Psi_{3}=z^{-1}\Psi_{3}^{(0)},\ \Psi_{4}=z^{-2}\Psi
_{4}^{(0)}{,} \label{try}%
\end{equation}
curvature invariants obviously do not depend on the choice of the reference
frame, $I=I^{(0)}$, $J=J^{(0)}$, and coefficients $K$, $L$, $N$ in certain
covariants (see Chapter 9.3 in \cite{kram}) transform according to
\begin{equation}
K=z^{-3}K^{(0)},\ L=z^{-2}L^{(0)},\ N=z^{-4}N^{(0)}.
\end{equation}
Here we use definitions (\ref{I}), (\ref{K}).

Usually, the parameter $z$ entering the boost (\ref{lna}) is finite and
non-vanishing, { so that classification criteria (whether eq.
(\ref{crit})\ are stisfied or not)} are not affected by the boost and all
timelike observers agree that the field belongs to the same type which is an
invariant characteristic of a spacetime at a given point. The situation is
qualitatively different on the horizon since $z\rightarrow0$ and thus, in
general, some of the quantities $K$, $L$, $N$, that vanish in the static frame
may or may not vanish in the freely falling one. This is obviously related to
the fact that the SO becomes null on the horizon and the corresponding null
frame is singular there. Consequently, only the results obtained in FFO's
frame should be considered as physically relevant.

For completeness, let us consider different cases separately, assuming that
all $\Psi^{\prime}s$ are finite on the horizon for both SO and FFO.
\\(1) $\Psi_{2}^{H}\neq0$. Then SO finds the gravitational field of type
D on the horizon, while FFO sees in general type II. Only if the metric
satisfies $K=N=0$ on the horizon, it is of type D also in FFO's frame.
\\(2) $\Psi_{2}^{H}=0$. Then SO sees type O on the horizon while FFO
sees type III, N or O depending on the behaviour of $\Psi_{3}$ and $\Psi_{4}$ there.

{ Thus, in general, there exists veriety of situations depending on the
relationship between invariants. It is convenient to summarize the set of
possible situations in the table:} \\[1mm]%

\begin{tabular}
[c]{|l|l|}\hline
SO & FFO\\\hline
D & II, D\\\hline
O & III, N, O\\\hline
\end{tabular}\\[1mm]

It is worth stressing that the static frame becomes singular on the horizon
and cannot be continued inside. This is one of the reasons why it is necessary
to introduce the Kruskal-like coordinate system near the horizon to obtain the
maximal analytical extension of the manifold. Note, e.g. that the fact that
all $\Psi^{\prime}s$ vanish on the horizon in the static frame (type O) does
not necessarily mean that the Weyl tensor as such vanishes - rather, this is a
pure coordinate effect because of a \textquotedblleft bad\textquotedblright%
\ frame and FFO (who uses a \textquotedblleft good\textquotedblright\ frame)
would see it in general non-vanishing. However, in a small vicinity of the
horizon the static frame is well-defined and, correspondingly, the horizon
limit has a clear sense. In  Ref. \cite{px}, where Petrov type on the
horizon for the axially symmetric case was considered, only the limit in the
SO sense was exploited. However, as was pointed out above, such a procedure is
not so \textquotedblleft innocent\textquotedblright\ and requires introducing
FFO for a complete and correct description. To conclude this section, let us
again emphasize that the horizon limit of Petrov type for SO and FFO in
general do not coincide.

\subsection{Examples - Ernst metric and Bonnor-Swaminarayan solution}

Let us here present two examples of exact solutions which are in general of
Petrov type I and become algebraically special on the event or acceleration horizon.

The Ernst metric \cite{ernst}
\begin{equation}
\fl ds^{2}=\Lambda^{2}\left[  -\left(  1-\frac{2m}{r}\right)  dt^{2}+\left(
1-\frac{2m}{r}\right)  ^{-1}dr^{2}+r^{2}d\theta^{2}\right]  +\Lambda^{-2}%
r^{2}\sin^{2}\theta d\phi^{2} {,}%
\end{equation}
where%
\begin{equation}
\Lambda=1+\frac{1}{4}B^{2}r^{2}\sin^{2}\theta{,}%
\end{equation}
represents a black hole immersed in an external magnetic field and for this
reason it is also called the Schwarzschild-Melvin spacetime in the literature.
The parameter $B$ governs strenth of the magnetic field. Let us choose the
frame (retaining the notation with prime of Ref. \cite{ernst}) 
\BEA
l^{\prime\mu}&=& \frac{1}{\sqrt{2}}\left(  \frac{1}{\Lambda}\chi^{-1/2}, \frac
{1}{\Lambda}\chi^{1/2},0,0 \right)  {,} \label{l'} \\
 n^{\prime\mu} &=&
\frac{1}{\sqrt{2}}\left(  -\frac{1}{\Lambda}\chi^{-1/2},\frac{1}{\Lambda}%
\chi^{1/2},0,0\right)  {, }\ \label{n'}\\ m_\mu&=&\frac{r}{\sqrt{2}}
\left(  0,0,-\Lambda,-\frac{i\sin\theta}{\Lambda}\right)  , \EEA
with $\chi=\left(  1-\frac{2m}{r}\right)  $, that corresponds to SO.

Then it follows that \cite{bose} 
\BEA
\Psi_0^\prime=C\chi=\Psi_4^\prime{, }\ \ C=\frac{3(\Lambda-1)(\Lambda
-2)}{r^2\Lambda^4} {,} \\
\Psi_1^\prime=-\Psi_3^\prime= C\chi^{1/2}
\cot\theta{, }\\\Psi_2^\prime=\frac{\Lambda-2}{r^2\Lambda^4}\left[
(2\cot^2\theta-1)(\Lambda-1)+\frac{m}{r}(3\Lambda-2)\right]  {,}\\
I=\chi^2C^2+4 C^2\chi\cot^2\theta+3\Psi_2^{\prime 2} {,} \\
 K^\prime=\chi^{3/2} K {, }\ \ K=3C^2\frac{\Lambda-2}{r^2\Lambda^4}k \cot\theta{,
}\\ 
k=\frac{m}{r}(3\Lambda-2) {,} \\ L^\prime=\chi L {,
}\ \ L=\frac{\Lambda-2}{r^2\Lambda^4}CA {, } \\ 
A=-\frac{\Lambda-1}%
{\sin^2\theta}+\frac{m}{r}(3\Lambda-2) {.} \\
 N^\prime=\chi^2N {,
}\ \ N=12L^2-C^2I {.}
\EEA
It is worth noting that there are no divergencies in $\Psi^{\prime}s$ and
quantities $K$, $N$, $L$ at $\theta=0$ and $\theta=\pi$ due to the factors $C$
and $\Lambda-1$ which are proportional to $\sin^{2}\theta$. In general,
$I^{3}\neq27J^{2}$ and the metric is of generic type I \cite{bose}. As was
pointed out in \cite{ortaggio}, the metric is of type II on the horizon. The
fact that it is algebraically special on the horizon can be easily seen since
$I^{3}-27J^{2}$ vanishes there. Since for the freely falling observer the
quantities $K$ and $N$ do not vanish on the horizon the Petrov type is II
there. In the static frame (\ref{l'}), (\ref{n'}), $K^{\prime}=L^{\prime
}=N^{\prime}=0$ on the horizon and this could lead to an incorrect conclusion
that the Petrov type on the horizon is D.

However, a special case arises on the axis of symmetry ($\theta=0$ or
$\theta=\pi$). It is seen from the above formulas that now all $\Psi^{\prime
}s$ vanish except $\Psi_{2}$ and both observers (a static and freely falling
one) will agree that the Petrov type is D. In the limit $B=0$ we recover the
Schwarzschild metric that is spherically symmetric and therefore we can rotate
a coordinate frame to achieve any point to lie on the axis. As a result, the
metric is of type D on the entire horizon that agrees with the fact that any
spherically symmetric metric is of the type D everywhere (\cite{w}, page 187).

Another special case arises if $\Lambda=2$. This can be achieved on the
horizon at $\sin\theta=(Bm)^{-1}$, if $Bm\geq1$. Then all $\Psi^{\prime}s$
vanish both for SO and FFO and thus the Weyl tensor is of type O.

Another interesting example of exact solutions with algebraically special Weyl
tensor on the horizon is the class of boost-rotation symmetric spacetimes
\cite{Bicak,Pravda}. These spacetimes correspond to various types of uniformly
accelerated ``particles'' and they thus possess an acceleration horizon. The
well known exact solution belonging to this class is the C-metric representing
two uniformly accelerated black holes that however does not suit our purposes
since it is globally of type D. Another interesting exact solution in this
class is the Bonnor-Swaminarayan solution \cite{Bonnor}, representing
uniformly accelerated Curzon-Chazy particles. This solution is in general of
Petrov type I and by evaluating $I^{3}-27J^{2}$ it can be checked that it
becomes algebraically special on the acceleration horizon.

\section{Truly naked black holes}

\label{sec4}

\subsection{Near-horizon behaviour of the Weyl scalars in the freely falling
frame}

If a null tetrad corresponds to the static frame (\ref{ln}), all Weyl scalars
are finite and, moreover, as we saw above, they vanish on the horizon, except
(possibly) $\Psi_{2}$. However, the appearance of the ``dangerous'' factor $z$
in the denominator opens the possibility that not only some Weyl scalar do not
vanish but that they may even diverge on the horizon in the freely falling
frame. To avoid such divergencies representing parallelly propagated curvature
singularity \cite{hawking}, some additional constraints should be imposed on
the metric. To distinguish metrics with finite and infinite $\Psi^{H\prime}s$,
let us now consider properties of the Weyl scalars in the freely falling frame
in more detail.

\subsubsection{Non-extremal case}

\label{tnbh-nec} Since in both frames $\Psi_{2}$ is the same and the horizon
values of $\Psi_{0}$ and $\Psi_{1}$ vanish, we need only to consider $\Psi
_{3}$ and $\Psi_{4}$. Then it follows from (\ref{01}), (\ref{c1}) and
(\ref{try}) that on the horizon%
\begin{equation}
\Psi_{3}^{H}=\frac{\bar{m}^{b}E}{\sqrt{2}\kappa_{H}}\left(  \frac{\kappa
_{2:b}}{2\kappa_{H}}+K_{;b}^{(1)}-\gamma^{cd}K_{bc;d}^{(1)}\right)
\end{equation}
is always finite.

Substituting the asymptotic expansion of the lapse function (\ref{nn}) and the
expansion%
\begin{equation}
K_{ab}=K_{ab}^{(1)}n+\frac{K_{ab}^{(2)}}{2}n^{2}+\frac{K_{ab}^{(3)}}{6}%
n^{3}+O(n^{4})
\end{equation}
in
\begin{equation}
\Psi_{4}=\frac{\bar{m}^{b}\bar{m}^{d}}{2}\exp(2\alpha)\left[  \frac{\partial
K_{bd}}{\partial n}-K_{bd}\frac{N^{\prime}}{N}+\left(  K^{2}\right)
_{bd}+\frac{N_{;b;d}}{N}\right]
\end{equation}
leads to
\begin{equation}
\Psi_{4}=2E^{2}\frac{\bar{m}^{a}\bar{m}^{b}}{\kappa_{H}^{2}}\left[
\frac{K_{ab}^{(2)}}{2n}+C_{ab}\right]  +O(n){,}%
\end{equation}
where
\begin{equation}
C_{ab}=\left[  K^{(1)2}\right]  _{ab}+\frac{\kappa_{2a;b}}{6\kappa_{H}}%
-\frac{\kappa_{2}}{3\kappa_{H}}K_{ab}^{(1)}+\frac{K_{ab}^{(3)}}{3}{.}%
\end{equation}
Thus, if we want $\Psi_{4}$ to be finite on the horizon we must demand
$K_{ab}^{(2)}=0$. Then $\Psi_{4}^{H}=2E^{2}\frac{\bar{m}^{a}\bar{m}^{b}%
}{\kappa_{H}^{2}}C_{ab}$. Note that for $K_{ab}^{(2)}=0$ all components of the
Riemann and Ricci tensors are also regular on the horizon.

\subsubsection{Extremal case}

Near the horizon%

\begin{equation}
\Psi_{3}=-\frac{\bar{m}^{b}E}{n_{0}\sqrt{2}}\frac{B_{;b}}{B^{2}}\exp\left(
\frac{n}{n_{0}}\right)  {.}%
\end{equation}
Thus, if we demand $\Psi_{3}$ to be finite on the horizon we must require
\begin{equation}
B_{;b}=0. \label{b0}%
\end{equation}
The limiting value of $\Psi_{4}$ on the horizon is
\begin{equation}
\Psi_{4}^{H}=2E^{2}\bar{m}^{b}\bar{m}^{d}\lim_{n\rightarrow\infty}\frac
{1}{N^{2}}\left[  \frac{\partial K_{bd}}{\partial n}-K_{bd}\frac{N^{\prime}%
}{N}+\left(  K^{2}\right)  _{bd}+\frac{N_{;b;d}}{N}\right]  {.}%
\end{equation}
Bearing in mind (\ref{b0}) we see that the most severe divergencies near the
horizon are absent because terms $\exp\left(  \frac{2n}{n_{0}}\right)  $ are absent.

To make sure that terms $\exp\left(  \frac{n}{n_{0}}\right)  $ are also
absent, we must pose a constraint on the correction to $N$, demanding that in
the expansion%

\begin{equation}
\fl N=\exp\left(  -\frac{n}{n_{0}}\right)  M {, }\ \ M=B+B_{1}\exp\left(
-\frac{n}{n_{0}}\right)  +B_{2}(x^{2},x^{3})\exp\left(  -\frac{2n}{n_{0}%
}\right)  +... \label{nexp}%
\end{equation}
not only $B$ but also $B_{1}$ do not depend on $x^{2}$ and $x^{3}$,
$B_{1;a}=0$. In contrast to the non-extremal case, finiteness of $\Psi_{4}$
does not entail constraints on $K_{ab}$.

\subsubsection{Ultraextremal case}

Now near the horizon $n\rightarrow\infty$, $z^{-1}=\frac{2E}{N}+O(N)$ and it
follows from (\ref{01}), (\ref{1}), (\ref{c1}), (\ref{0}), (\ref{try}) and
(\ref{ue}) - (\ref{kabu}) that
\begin{equation}
\Psi_{3}^{H}=\frac{\bar{m}^{b}E}{\sqrt{2}}\lim_{n\rightarrow\infty}\frac{1}%
{N}\left(  N^{-1}\frac{\partial N_{;b}}{\partial n}+\frac{K_{b}^{c}N_{;c}}%
{N}+K_{;b}-\gamma^{cd}K_{bc;d}\right)  {,}%
\end{equation}%
\begin{equation}
\Psi_{3}=-\frac{\bar{m}^{b}E}{\sqrt{2}}\frac{n^{m-1}}{A^{2}}[mA_{;b}%
+O(n^{-1})]{,}%
\end{equation}%
\begin{equation}
\Psi_{4}=\frac{2E^{2}}{A^{2}}\bar{m}^{b}\bar{m}^{d}n^{2m}[\frac{A_{;b;d}}%
{A}+O(n^{-1})]{.}%
\end{equation}
The condition
\begin{equation}
A_{;b}=0\ \label{a0}%
\end{equation}
is necessary to ensure the finiteness of $\Psi_{3}$ and $\Psi_{4}$ on the horizon.

Let us point out that the conditions of constancy of $B$ or $A$ on the
(ultra)extremal horizon also ensure that in the static frame $\Psi_{3}$ and
$\Psi_{4}$ vanish and thus, similarly as in the non-extremal case, the field
from SO's viewpoint is of type D or O.

\subsection{Higher order curvature invariants}

From the beginning we assumed that the Kretschmann invariant as well as other
polynomial invariants of the Riemann tensor are regular on the horizon as was
done in \cite{med}. In the previous section we showed that even if this
regularity condition is satisfied, $\Psi_{4}$ may diverge in the horizon limit
in the parallelly propagated frame attached to FFO. Here we show that even
when all components of the Weyl tensor in the paralelly propagated frame are
regular, curvature invariants constructed from second derivatives of the
Riemann tensor may still diverge on the horizon.

First let us point out that if $\Psi_{4}$ is finite on the horizon, then it
turns out that the first order\footnote{A curvature invariant is said to be of
order $n$ if it contains covariant derivatives of the Riemann tensor up to
$n$.} curvature invariant $I_{d1} =R_{\alpha\beta\gamma\delta;\epsilon
}R^{\alpha\beta\gamma\delta;\epsilon}$ is regular on the
horizon\footnote{Recently, certain constraints on the expansion (\ref{nn}),
(\ref{expgamma}) of the metric in the vicinity of the horizon were obtained in
\cite{cvitan} by demanding regularity of a first order curvature invariant
instead of regularity of the components of the Weyl tensor in a paralelly
propagated frame.}.

In order to express the second order curvature invariant $I_{d2}%
=R_{\alpha\beta\gamma\delta;\epsilon\eta}R^{\alpha\beta\gamma\delta
;\epsilon\eta}$ we invoke the remaining coordinate freedom and substitute
$[\gamma_{H}]_{ab}(x,y)=\exp{\left(  2\theta(x,y)\right)  }\delta_{ab}$ in the
expansion (\ref{expgamma}) \cite{med}. Then we obtain
\begin{equation}
\fl I_{d2}=\left(  \frac{9}{16}\exp\left(  -4\theta(x,y)\right)
\mbox{Sp}\left(  [\gamma_{3}](x,y)^{2}\right)  +\frac{1}{4{\kappa_{H}}^{2}%
}{\kappa_{3}(x,y)}^{2}\right)  n^{-2}+O(n^{-1}),
\end{equation}
where $\mbox{Sp}([\gamma_{3}]^{2})=([\gamma_{3}]_{11})^{2}+2([\gamma_{3}%
]_{12})^{2}+([\gamma_{3}]_{22})^{2}$. The term proportional to $n^{-1}$ is
quite complicated, but the only information we need is that it vanishes for
$\mbox{Sp}([\gamma_{3}]^{2})=0=\kappa_{3}$. According to Sec. \ref{tnbh-nec},
the first term $\mbox{Sp}([\gamma_{3}]^{2})$ in the invariant $I_{d2}$ has to
vanish if we want to avoid parallelly propagated curvature singularity. It is
worthwhile to note that $[\gamma_{3}]\sim K_{ab}^{(2)}$. As is shown in
preceding subsections, this quantity vanishes if all Weyl scalars are finite
on the horizon. However, for $\kappa_{3}(x,y)\not =0$ the invariant $I_{d2}$
still diverges on the horizon. The case with diverging higher order curvature
invariants is often also considered as a curvature singularity (see e.g.
\cite{w}) but we see no reason why timelike geodesics could not be extended
through the horizon since all zeroth and first order curvature invariants are
regular on the horizon. Furthermore, either frame components of the Weyl,
Ricci and Riemann tensors as well as the energy density measured by FFO in the
parallelly propagated frame are also finite there or we are faced with
a new type of a horizon which is considered in the next subsection and that we
now turn to (see also Sec. \ref{EinstT}).

\subsection{Regularity}

We have seen that finiteness of the Kretschmann invariant $R_{\alpha
\beta\gamma\delta}R^{\alpha\beta\gamma\delta}$ does not guarantee by itself
finiteness of all curvature components since different terms can enter this
expression with different signs due to the Lorentz signature. This does not
happen in the static frame where all non-vanishing contributions to the
Kretschmann invariant have the same sign but it occurs in the freely falling
frame where some divergencies in particular components can be mutually
canceled. The same is true with respect to higher order curvature invariants.
To gain insight, it is instructive to consider the simplest spherically
symmetric case where the metric can be cast in the form
\begin{equation}
ds^{2}=-\frac{F}{G}dt^{2}+\frac{d\rho^{2}}{F}+r^{2}(\rho)(d\theta^{2}+\sin
^{2}\theta d\varphi^{2}) {.}%
\end{equation}

It was observed in \cite{nk1}, \cite{nk2} that the transition from a static
frame to a freely falling one leads in general to significant enhancement of
some curvature components due to the divergent factors like $\cosh\alpha.$ In
particular, in the orthonormal static frame
\begin{equation}
R_{\hat{0}\hat{2}\hat{0}\hat{2}}^{(0)}+R_{\hat{1}\hat{2}\hat{1}\hat{2}}%
^{(0)}=\frac{F}{r}\left[  -r^{^{\prime\prime}}(\rho)-\frac{G^{\prime}}%
{2G}r^{\prime}(\rho)\right]  {,} \label{012}%
\end{equation}
while in the freely falling frame
\begin{equation}
R_{\hat{1}\hat{2}\hat{1}\hat{2}}+R_{\hat{0}\hat{2}\hat{0}\hat{2}}=(2\cosh
^{2}\alpha-1)\left(  R_{\hat{1}\hat{2}\hat{1}\hat{2}}^{(0)}+R_{\hat{0}\hat
{2}\hat{0}\hat{2}}^{(0)}\right)  {.}%
\end{equation}

As one approaches the horizon $F(\rho_{+})=0$, the factor cosh$\alpha$
diverges but the combination (\ref{012}) tends to zero and, as a result of the
competition of these two factors, $R_{\hat{1}\hat{2}\hat{1}\hat{2}}+R_{\hat
{0}\hat{2}\hat{0}\hat{2}}$ remains finite (see, e.g., Eq. (2.12) of Ref.
\cite{nk1}). This is, however, not necessarily the case for the distorted
horizon. In turn, divergent components of the Riemann curvature tensor lead to
divergencies in some Weyl components. In particular, for the non-extremal
horizon we have seen that the component $\Psi_{4}$ remains finite only
provided the terms of the order $n^{2}$ in $K_{ab}$ vanish near the horizon
(equivalently, the terms of the order $n^{3}$ vanish in $\gamma_{ab}$). In a
spherically symmetric spacetime such a condition simply follows from the
analyticity of $F(\rho)$, $G(\rho)$ and $r(\rho)$ since the expansion in
$\rho-\rho_{+}$ is equivalent to the expansion in even powers of $n$.
Alternatively, for spherically symmetric spacetimes all $\Psi$'s except
$\Psi_{2}$ vanish everywhere, so that the transformation (\ref{try}) does not
change this circumstance irrespective of the fact that on the horizon $z$
vanishes. This is simply manifestation of the known fact that all spherically
symmetric spacetimes belong to type D \cite{w}.

Thus, as a matter of fact, \textquotedblleft naked black
holes\textquotedblright\ \cite{nk1}, \cite{nk2} are not naked in the sense
that in both frames curvature components are finite. Meanwhile, for a
distorted horizon a new possibility opens when curvature components are finite
and non-zero in the static frame but some of them become infinite in the
freely falling one. As the term \textquotedblleft naked black
holes\textquotedblright\ is already reserved, such objects can be called
\textquotedblleft truly naked black holes\textquotedblright\ (TNBH). We cannot
indicate the concrete examples of such objects because of obvious difficulties
connected with finding exact solutions without spherical symmetry but, in any
case, we see no reason why they should be rejected in advance. Apart from
this, it is just the condition for absence of TNBH that enables us to
understand better the conditions for the curvature components that should hold
on the horizon to make it perfectly regular in any frame. It is natural to
call it \textquotedblleft strong regularity\textquotedblright\ of the horizon.
In a similar way, one can introduce the condition of regularity of invariants
$I_{dn}$ with n-th derivatives of the Riemann or Weyl tensor. As was stressed
in \cite{brill95}, establishing properties of a generic spacetime connected
with analyticity and/or regularity is not an easy task. However, using the
conditions of absence of \ TNBH and regularity of $I_{dn}$, we can suggest (by
analogy with the spherically-symmetrical case) the generalization of notion of
regularity applicable to distorted spacetimes: if all curvature invariants
composed from the curvature components and their derivatives up to the order
$m$ are finite on the horizon, the metric can be said to be regular up to the
order $m$.

\subsection{Structure of Einstein and stress-energy tensors on the horizon}

\label{EinstT}

The typical feature of TNBH is that the structure of the Einstein tensor on
the horizon changes as compared with usual black holes. Before discussing this
point, we would like to comment on some generic properties of the horizon. It
was observed in \cite{med} that on the static regular horizon the Einstein
tensor should obey the relationships%
\begin{equation}
G_{1}^{1}=G_{0}^{0}{,}\ \ G_{1a}=0{} \label{bl}%
\end{equation}
and (if Einstein's equations are fulfilled) the similar ones for the
stress-energy tensor $T_{\mu}^{\nu}$. Meanwhile, these equalities immediately
follow from the general conditions on the non-extremal horizon%
\begin{equation}
R_{\alpha\beta}l^{\alpha}l^{\beta}=0 \label{rllhor}%
\end{equation}
and%
\begin{equation}
R_{\alpha\beta}l^{\alpha}m^{\beta}=0 \label{rlmhor}%
\end{equation}
(see. e.g. \cite{nf}, Eqs. (6.2.2) and an unnumbered equation after Eq.
(6.3.29)). Indeed, substituting (\ref{ln}) into (\ref{rllhor}), (\ref{rlmhor}%
), we immediately arrive at (\ref{bl}). On the other hand, the advantage of
proof of equations (\ref{bl}) in \cite{med} is that it does not use the weak
and dominant energy conditions, on which Eqs. (\ref{rllhor}), (\ref{rlmhor})
usually rely \cite{nf}, and arises as a pure geometrical property. (We recall
that the constancy of the surface gravity for static Killing horizons was also
demonstrated in \cite{med} without using the energy conditions.)

Note that the properties (\ref{bl}) were established for the non-extremal
horizons only. Let us see what happens in the case of (ultra)extremal
horizons. It follows from Eqs. (40), (42) - (44) of Ref. \cite{med} that
\BEA
\ \ \ \ \ G_{1a} &=& K_{;a}-K_{ab;c} \gamma^{cb}-K_a^b\frac{N_{;b}}{N}-\frac
{\partial_nN_{;a}}{N} {,}\\ G_1^1-G_0^0 &=& -{\it Sp}K^2-K\frac
{N^\prime}{N}+K^\prime+\frac{\Delta_2N}{N} {.} \label{11-00} \EEA

On the horizon all terms with $K$ vanish. With the asymptotic form of the
lapse function (\ref{e}) one obtains that on the extremal horizon 
\BEA
\ \ \ \ \ \ \ G_{1a}^H &=& \frac{B_{;a}}{n_{0B}} {,} \label{1a}\\%
(G_1^1-G_0^0)^H &=& \frac{\Delta_2B}{B} {.} \label{g01} \EEA
In a similar way we obtain for the ultraextremal case (\ref{ue}) that
 \BEA
\ \ \ \ \ \ \ G_{1a}^H &=& 0 {,} \label{ue1}\\(G_1^1-G_0^0)^H &=&
\frac{\Delta_2A}{A} {.} \label{g01a} \EEA

Thus, the block-diagonal structure of $G_{\mu}^{\nu}$ and $T_{\mu}^{\nu}$,
typical for non-extremal horizons, fails for an extremal horizon and the
``equation of state'' $p_{\mid\mid}\equiv-\rho$\ ($p_{\mid\mid}\equiv
T_{1}^{1}$ is a longitudal pressure, $\rho=-T_{0}^{0}$ is the energy density)
does not hold in that case. Moreover, additional stresses $T_{1a}$ appear on
the horizon. For the ultraextremal horizon, the block-diagonal structure
retains its validity as it is seen from (\ref{ue1}) but the horizon ``equation
of state'' changes. It is worth stressing that it is the combined effect of
non-sphericity and (ultra)extremality that leads to such changes. For
spherically symmetric spacetimes $\Delta_{2}A=\Delta_{2}B=0$ and the horizon
structure of $G_{\mu}^{\nu}$ and $T_{\mu}^{\nu}$ coincides with that of
non-extremal horizons.

Note that, as $\int d^{2}x\sqrt{\gamma}\Delta_{2}A=$ $\int d^{2}x\sqrt{\gamma
}\Delta_{2}B=0$ over a closed surface, these quantities should change a sign
somewhere on the cross-section of the horizon \mbox{$(t={\rm const}$,}
$n\rightarrow\infty)$. It means that $\xi\equiv(p_{\mid\mid}+\rho)^{H}$ also
changes a sign, so that the null energy condition ($T_{\mu\nu}l^{\mu}l^{\nu
}\geq0$ for any null vector) is violated in some region on the horizon
surface. In other words, there are regions on the (ultra)extremal distorted
horizon where $\xi<0$, so that the matter source becomes ``phantomic''. In
recent years, such a type of source has been discussed intensively in
cosmology (see, e.g., \cite{cald}) but we see that it arises in the black hole
context as well.

The relationships (\ref{1a}), (\ref{g01}) - (\ref{g01a}) are obtained from the
regularity conditions in the static frame only. If, additionally, we assume
that the horizon is not ``truly naked'' then the equalities (\ref{b0}),
(\ref{a0}) should hold. As a consequence, all extra terms in (\ref{1a}),
(\ref{g01}) - (\ref{g01a}) vanish and we return to the ``normal'' relations
(\ref{bl}) so that there is no difference in this point between the
non-extremal and (ultra)extremal horizons. In a similar way, the extra terms
vanish in $\Psi_{2}^{H}$ in (\ref{2horb}), (\ref{2hor}).

One can also carry out the following analogy between non-extremal and
(ultra)extremal horizons. In the first case the leading term of the
asymptotics of the lapse function $N$ is determined by the surface gravity
$\kappa_{H}$ according to (\ref{nn}). The demand of regularity of the
Kretschmann invariant on the horizon leads to the constancy of the surface
gravity there: $\kappa_{H;a}=0$. In the second case $\kappa_{H}=0$ and the
main term in the asymptotics is determined by the coefficient $B$ (or $A$). If
we demand the regularity of not only the Kretschmann scalar but also of all
curvature components in the freely falling frame, we obtain the similar
condition $B_{;a}=0$ (or $A_{;a}=0$). In this sense, the condition of absence
of truly naked extremal horizons looks formally like the analog of the zeroth
law of mechanics of non-extremal black holes.

The condition of finiteness of all components of the Riemann (or Weyl) tensors
on the horizon in the freely falling frame can be also understood as follows.
The energy measured by a freely falling observer is equal to%
\begin{equation}
\fl\varepsilon=T_{\mu\nu}u^{\mu}u^{\nu}=\cosh^{2}\alpha\varepsilon^{(0)}%
+\sinh^{2}\alpha p_{\mid\mid}^{(0)}=\frac{E^{2}}{N^{2}}\left[  \varepsilon
^{(0)}+p_{\mid\mid}^{(0)}\right]  -p_{\mid\mid}^{(0)}{,} \label{tuu}%
\end{equation}
where we used (\ref{ua}), (\ref{ea}) and (\ref{ues}). It follows from the
Einstein equations $\varepsilon^{(0)}+p_{\mid\mid}^{(0)}=\frac{G_{1}^{1}%
-G_{0}^{0}}{8\pi}$, where the later quantity is given by Eq. (\ref{11-00}).
Meanwhile, the Ricci component in the FFO's frame $R_{\alpha\beta}n^{\alpha
}n^{\beta}=\frac{G_{1}^{1}-G_{0}^{0}}{2z^{2}}$, where near the horizon
$z^{-2}=\frac{4E^{2}}{N^{2}}$. Thus, near the horizon eq. (\ref{tuu}) can be
rewritten also as $\varepsilon=\frac{R_{\alpha\beta}n^{\alpha}n^{\beta}}%
{16\pi}-p_{\mid\mid}^{(0)}$.

Consider the non-extremal case. In the limit $n\rightarrow0$ we obtain that
\mbox{$\varepsilon=\frac{E^{2}K_{a}^{a(2)}}{16 \pi \kappa_{H}^{2}}n^{-1}+O(1)$}
so that $\varepsilon$ diverges on the horizon unless $K_{a}^{a(2)}=0$. Thus,
the condition $K_{ab}^{(2)}=0$ necessary for the finiteness of $\Psi_{4}$
ensures also the finiteness of $\varepsilon$ on the horizon.

Consider now the extremal case and substitute the expansion (\ref{nexp}) into
(\ref{tuu}). Then near the horizon
\begin{equation}
\varepsilon=\frac{E^{2}}{8\pi M^{3}}\left[  \Delta B\exp\left(  2\frac
{n}{n_{0}}\right)  +\Delta B_{1}\exp\left(  \frac{n}{n_{0}}\right)  +\Delta
B_{2}\right]  +...
\end{equation}
and the conditions $B_{;a}=0=B_{1;a}$ ensuring the finiteness of $\Psi_{4}$
ensure also the absence of singular terms in $\varepsilon$. The similar result
takes place for ultraextremal black holes. In other words, in all three cases
either we have a regular horizon without TNBH and finite $\varepsilon$ or TNBH
and infinite $\varepsilon$.

\section{Ricci tensor}

\label{sec5}

There are several methods of classification of the Ricci tensor (see e.g.
\cite{pleb,penrind,joly,hall,kram}). One possible method is to construct so
called Pleba\'{n}ski tensor, with same symmetries as the Weyl tensor, from the
traceless part of the Ricci tensor, ${S}_{\alpha\beta}=R_{\alpha\beta}%
-\frac{R}{4}g_{\alpha\beta}$. Then one can classify the Ricci tensor according
to the Petrov type of the Pleba\'{n}ski tensor. Corresponding algebraic type
is then called Petrov-Pleba\'{n}ski (PP) type. Another more detailed
classification is the Segre classification based on geometric multiplicities
of eigenvalues of ${S}_{\alpha\beta}$.

For determining to which class the Ricci tensor on the horizon belongs, we
will use the classification algorithm and notation given in \cite{joly}. We
will thus need to express curvature invariants $I_{6}$, $I_{7}$ and $I_{8}$
and analyze whether certain syzygies between them are satisfied.

First step is to determine whether
\begin{equation}
Q\equiv{I_{R}}^{3}-27{J_{R}}^{2}=0 {, }%
\end{equation}
where\BEA
I_R&=&\frac{1}{48}\left(  7{I_6}^2-12{I_8}\right)  {,}\\ J_R&=&\frac
{1}{1728}\left(  36I_6I_8-17{I_6}^3-12{I_7}^2\right)  {,}\\
I_6&=&S_\mu^\nu S_\nu^\mu{, }\ \ I_7=S_\mu^\nu S_\alpha^\mu S_\nu^\alpha{,
}\ \ I_8=S_\mu^\nu S_\alpha^\mu S_\beta^\alpha S_\nu^\beta{.}\EEA

Using the formulas of 2+1+1 decomposition {\it (\ref{00}) - (\ref{ab})}, it
is straightforward to find that 
\BEA 
S_0^0&=&\frac{1}{4}\left(  { Sp} K^2+K^2-R_\parallel\right)  +\frac
{1}{2}\left(  K\frac{N^\prime}{N}-\frac{N^{\prime\prime}}{N}-K^\prime
-\frac{\Delta_2N}{N}\right)  ,\\
 S_1^1&=&\frac{K^\prime}{2}-\frac{3}%
{4}{ Sp} K^2-\frac{1}{2}\frac{N^{\prime\prime}}{N}-\frac{1}{4}\left(
R_\parallel-K^2-\frac{2\Delta_2N}{N}+2K\frac{N^\prime}{N}\right)  ,\\
S_{ab}&=&\frac{\partial K_{ab}}{\partial n}+2(K^2)_{ab}-K_{ab} K+\frac{1}{N}\left(
K_{ab} N^\prime-N_{;ab}\right)  +\gamma_{ab}\frac{L}{4} {,}\\
 L&=&R_\parallel
-R_\perp-2K^\prime+ { Sp} K^2+K^2+\frac{2\Delta_2N}{N}-2K\frac{N^\prime
}{N},\\
 S_{1a}&=&R_{1a}=K_{;a}-K_{ab;c}\gamma^{bc}-K_a^b\frac{N_{;b}}{N}%
-\frac{\partial_nN_{;a}}{N} {.}
\EEA

On the non-extremal horizon 
\BEA
S_{1a}^H&=&0 {,} \label{s1a}\\
 S_1^{1(H)}&=&S_0^0(H)\equiv-\alpha{,}
\label{al}\\ S_{ab}^H&=&\gamma_{ab} \alpha+\mu_{ab} , \ \ \mu_{ab} \equiv
2K_{ab}^{(1)}-\gamma_{ab} K^{(1)} , \ \ { Sp}\hat{\mu}=\mu_a^a=0 {.} 
\EEA

It is worth noting that, for our choice (\ref{ln}), Eq. (\ref{al}) is
equivalent to (\ref{rllhor}) and Eq. (\ref{s1a}) is equivalent to
(\ref{rlmhor}).

The expression for 2x2 matrix $S_{ab}$ can be rewritten in the form%
\begin{equation}
\hat{S}=\alpha\hat{I}+\hat{\mu} {.}%
\end{equation}
It is convenient to represent the traceless symmetrical 2x2 $\hat{\mu}$ matrix
in the form%
\begin{equation}
\hat{\mu}=a\sigma_{z}+b\sigma_{x} {,}%
\end{equation}
where $\sigma_{z}$ and $\sigma_{x}$ are Pauli matrices and $a=\mu_{11}$,
$b=\mu_{12}$. Then, using the properties of Pauli matrices $\sigma_{x}%
^{2}=\sigma_{z}^{2}=1$ and $\mathrm{Sp}\sigma_{x}=\mathrm{Sp}\sigma_{z}=0$, it
is easy to find that\BEA
I_6&=&2(2\alpha^2+\beta^2) , \ \ \beta^2\equiv a^2+b^2 {,}\\
I_7&=&6\alpha\beta^2 {,}\\ I_8&=&2(2\alpha^4+6\alpha^2\beta
^2+\beta^4),\\ I_R&=&\frac{A}{48} , \ \ A=4(4\alpha^2-\beta^2)^2
{,}\\ J_R&=&\frac{(-64\alpha^6+48\alpha^4\beta^2-12\alpha^2\beta
^4+\beta^6)}{216} \EEA
and direct comparison shows that, indeed, $Q=0$. Since $I_{7}$ is in general
non-vanishing, we can conclude that Ricci tensor has in the generic case one
pair of equal eigenvalues and thus list of its algebraic multiplicities is
$\{112\}$. Corresponding Segre types are $[112]$ (PP-type II) and $[(11)1,1]$,
$[11(1,1)]$, and $[(11)Z\bar Z]$ (PP-types D).

Similarly, the same result may be obtained for (ultra)extremal horizons.

Note also that, if the TNBH is excluded, the Ricci components
\mbox{$\Phi_{00}=\frac{1}{2} {S}_{\alpha\beta} l^{\alpha} l^{\beta}$} and
\mbox{$\Phi_{01} = \frac{1}{2} {S}_{\alpha\beta} l^{\alpha} m^{\beta}$} vanish
on the horizon both in SO's and FFO's frame (this immediately implies that the
PP-type is algebraically special, but we still need invariant $I_{7}$ to rule
out more special cases).

Now, let us say that a quantity $x$ which under a boost ${l^{\prime}}^{\alpha
}=zl^{\alpha}$, ${n^{\prime}}^{\alpha}=z^{-1}n^{\alpha}$ changes according to
$x^{\prime}=z^{q}x$ has a boost weight $q$. We may conclude that on the
horizon all components of the Weyl, Ricci and also Riemann and Einstein
tensors with positive boost weight vanish. This implies that all these tensors
are aligned\footnote{i.e. they have common aligned null direction defined in
\cite{milson}}, algebraically special on the horizon in the sense of
\cite{milson} and of the alignment type (2).

\section{Summary and conclusions}

\label{sec6}

We have analyzed various properties of the curvature tensors in the vicinity
of a generic static Killing horizon.

For both SO and FFO the Weyl, Ricci, Riemann and Einstein tensors on the
horizon are algebraically special of alignment type (2) with a common aligned
null direction. However, for FFO and SO the horizon limits in general do not
coincide since SO becomes null and the corresponding frame is singular on the
horizon. Consequently, only the results obtained in the freely falling frame
should be regarded as relevant.

It turns out that the horizon of a generic static black hole is in general of
Petrov type II. More special types (D,III,N,O) are also possible.
{ Further details depend on relationships between Petrov invariants that
do not follow directly from the properties of the Killing horizon and,
therefore, cannot be determined in a general form. }

Possible Segre types of the Ricci tensor on the horizon are $[112]$,
$[(11)1,1]$, $[11(1,1)]$, and $[(11)Z\bar{Z}]$ and more special.

It is found that the notion of regularity on the horizon requires more subtle
definition. Due to the Lorentz signature, finiteness of the Kretschmann
invariant on the horizon does not exclude divergencies in some components of
the Weyl tensor in the freely falling frame or, equivalently, infinite tidal
forces or energy density as measured by FFO. It turns out that the horizon may
look regular from the viewpoint of SO but singular from the viewpoint of FFO
and the conditions (``strong regularity'' of a horizon) that exclude such
exotic objects (``truly naked black holes'' - TNBH) are given. In doing so,
for (ultra)extremal case we obtained a formal analog of the zeroth law of
mechanics of non-extremal black holes.

It has also turned out that for non-extremal horizons it is necessary to apply
an additional restriction on expansion of the lapse function together with
strong regularity conditions in order to guarantee that a curvature invariant
composed from the second derivatives of the Riemann tensor is also finite.

It is shown that the structure of a stress-energy tensor for distorted truly
naked horizons differs from the non-extremal case, so that the block-diagonal
structure of the stress-energy tensor fails. In doing so, the stress-energy
tensor should somewhere on the horizon have a phantomic-like equation of
state. This reveals itself for non-spherical horizons only as a combined
effect of non-sphericity, extremality and presence of infinite tidal forces
for FFOs. Once we impose the condition of strong regularity, this effect
immediately disappears.

\ack
O.Z. would like to thank Mathematical Institute of the Czech Academy of
Sciences for its hospitality while part of this work was carried out. V.P. is
grateful to M. Ortaggio for discussions and for pointing out reference
\cite{px} and to A. Pravdov\'{a} for discussions. V.P. was supported by
institutional research plan No. AV0Z10190503 and by grant KJB1019403.

\end{document}